\documentclass[a4paper,11pt]{article}
\usepackage{jcappub} 

\usepackage{graphicx}
\usepackage{caption}
\usepackage{subcaption}
\usepackage{mathrsfs}

\usepackage{siunitx}
\let\svqty\qty
\usepackage{physics}
\let\qty\svqty
\usepackage{amsmath,amssymb,amsfonts}
\def\be#1\ee{\begin{align}#1\end{align}} 
\def\bse#1\ese{\begin{subequations}#1\end{subequations}}

\newcommand{\kn}{\ensuremath{\kappa_\text{n}}} 
\newcommand{\ki}{\ensuremath{\kappa_\text{i}}} 
\newcommand{\kp}{\ensuremath{\kappa_\text{p}}} 

\usepackage[capitalize]{cleveref}
\crefname{equation}{eq.}{eqs.}
\crefname{figure}{figure}{figures}
\crefname{table}{table}{tables}
\crefname{subequation}{eqs.}{eqs.}
\labelcrefformat{subequation}{#2(#1)#3}
\crefformat{subequations}{#2(#1)#3}
\crefname{section}{section}{sections}
\crefname{appendix}{appendix}{appendices}

\usepackage{comment}
\usepackage{soul}

\newcommand{\qu}[1]{``#1''} 

\author[a,b]{Francesco Del Porro,}
\affiliation[a]{Center of Gravity, Niels Bohr Institute, Blegdamsvej 17, 2100 Copenhagen, Denmark}
\affiliation[b]{Niels Bohr International Academy, Niels Bohr Institute, Blegdamsvej 17, 2100 Copenhagen, Denmark}
\emailAdd{francesco.del.porro@nbi.ku.dk}

\author[c]{and Jacopo Mazza}
\affiliation[c]{Université Paris-Saclay, CNRS/IN2P3, IJCLab, 91405 Orsay, France}
\emailAdd{jacopo.mazza@ijclab.in2p3.fr}

\title{\boldmath Mechanics of non-Killing horizons}
\abstract{We investigate the mechanics of stationary axisymmetric non-Killing horizons, which emerge in spacetimes that do not enjoy the symmetry known as circularity --- as is commonly the case for rotating black holes beyond general relativity. 
Specifically, we define and compute three notions of surface gravity: inaffinity, normal, and peeling; and find that the inaffinity and normal definitions generically differ, while the normal and peeling definitions always agree, although none of them is constant over the horizon. We then derive a version of Smarr's formula, which appears to involve an average over the horizon of the normal surface gravity. We also compute, via the tunnelling method, the spectrum of Hawking's radiation, verifying that its temperature is controlled by the (non-constant) peeling surface gravity. Finally, we recapitulate the status of the four laws of black hole mechanics in situations in which the event horizon fails to be Killing.
Our results thus pave the way to a deeper understanding of black hole thermodynamics beyond general relativity.
}

\begin{document}
\maketitle
\flushbottom

\section{\label{sec:intro}Introduction}

The development of black hole thermodynamics has been a major advancement in theoretical physics, since it
\qu{has brought to light strong hints of a very deep and fundamental relationship between gravitation, thermodynamics, and quantum theory} --- the quote is from  \cite{WaldThermodynamicsBlack2001}.
This line of research is based on two fundamental building blocks: 
the identification of an analogy connecting certain laws of \qu{horizon mechanics} to the usual laws of thermodynamics \cite{BardeenFourLaws1973};
and the realisation that the canonical quantisation of fields on a background containing a horizon entails that an observer at infinity detects the production of particles populating a thermal bath, whose temperature is set by local properties of the horizon --- a phenomenon known as Hawking radiation \cite{Hawking_1974_BHexplosion,HawkingParticleCreation1975}. 
Reviews on the subject abound and include, for instance, \cite{WaldQuantumField1995,WaldThermodynamicsBlack2001,Unruh:1976db,CarlipBlackHole2014,KubiznakBlackHole2017,MannBlackHole2025}.

In vacuum general relativity, specific properties of the stationary and axisymmetric solutions to the field equations ensure that the quantity playing the role of the temperature in the analogy with the laws of thermodynamics --- i.e.~the surface gravity --- is a constant, in the sense that it is the same at each point on a black hole's horizon.
Moreover, one can prove that this temperature is the same as that of the thermal bath of Hawking particles. 
These facts have corroborated the interpretation of such solutions as thermal equilibrium states, 
with an internal energy associated to quantities that are conserved in physical processes, and with an entropy proportional to the horizon's area \cite{BekensteinBlackHoles1973,HawkingBlackHoles1976}.

Technically, these properties descend from the fact that, in vacuum general relativity, the event horizon of a stationary and axisymmetric solution of the field equations is also a \emph{Killing horizon}.  
These black holes are therefore highly symmetric, a fact that is not at all trivial and constitutes the core of the so-called rigidity theorem \cite{HawkingLargeScale1973}.
The field equations of general relativity thus play a rather important role in the derivation of these results, and consequently in the interpretation of black holes as thermodynamic systems.

Crucially, however, black holes beyond general relativity need not enjoy the same properties, as their event horizon is not guaranteed to be a Killing horizon.
(The same is true, to some extent, for non-vacuum black holes in general relativity, for that matter.)
Black holes without Killing horizons are not necessarily \qu{exotic}, as often they play the same role in well-defined modified theories of gravity as the Kerr solution does in general relativity.
Simply put, gravity beyond general relativity is expected to be \qu{more complicated} than what general relativity itself describes; hence, the fact that black holes reflect such an increased level of complexity is not surprising.
In more technical terms, the non-Killing nature of the horizon can be traced back to the breaking of a symmetry known as \emph{circularity}, which can be proven to hold in general relativity and several other contexts, but cannot be assumed in completely general situations.

The goal of this article, therefore, is to extend familiar notions related to horizon mechanics to the context of non-Killing horizons.
Our treatment will echo that of previous works on the subject, e.g.~\cite{AshtekarIsolatedHorizons1999,AshtekarGenericIsolated2000,AshtekarMechanicsIsolated2000,AshtekarMechanicsRotating2001,AshtekarDynamicalHorizons2003,HaywardGeneralLaws1994,NielsenBlackHoles2009,LiHawkingRadiation2021}, but is substantially novel in its focus on the breaking of circularity.
Specifically, \cref{sec:circ} will serve as an introduction to non-Killing horizons and to their relationship with the breaking of circularity; incidentally, we will show that if circularity holds on the horizon, then the horizon must be Killing.
\Cref{sec:surfgrav} will be dedicated to the notion of surface gravity, in three of its realisations: normal, inaffinity, and peeling; in particular, we will highlight the relationships among the different definitions, and compute explicit expressions in terms of the components of the metric.
\Cref{sec:thermo} will be an invitation towards the thermodynamics of non-Killing horizons: we will derive a version of Smarr's formula, discussing the differences with respect to the usual case; and show that non-Killing horizons emit Hawking radiation in a way that is essentially analogous to the familiar case, except for the fact that the temperature of said radiation is not constant, but it depends on the location on the horizon at which the Hawking quantum is emitted; finally, we will collect some remarks on the four laws of black hole mechanics and on their thermodynamical interpretation.
\Cref{sec:concl} closes the article with a critical discussion of our results.

\section{\label{sec:circ}Non-Killing horizons and the lack of circularity}

We focus on the exterior region of stationary axisymmetric and asymptotically flat black holes.
We call $\xi^\mu$ the Killing vector associated to time translations, and assume it to be timelike at least in a neighbourhood of infinity.
We further call $\psi^\mu$ the Killing vector that generates rotations around an axis, and assume it to be spacelike, with closed orbits, and vanishing on the axis of symmetry.
A classical result due to Carter ensures that such Killing vectors can be taken to commute \cite{CarterCommutationProperty1970}.

When referring to the horizon of said black holes, we always have in mind the \emph{future event horizon}, defined as the boundary of the causal past of future null infinity.
In many situations, e.g.~in vacuum general relativity, such horizon is also a \emph{Killing horizon}, meaning that there exists a Killing vector normal to it.
As we will discuss at length in this article, this coincidence has profound consequences on the thermodynamic interpretation of black holes.

However, examples of black holes without Killing horizons exist.
Probably the neatest ones are the so-called disformed Kerr solution in scalar--tensor gravity \cite{AnsonDisformingKerr2021,BenAchour:2020fgy}, and the other disformed solutions in scalar--tensor and vector--tensor theories obtained in \cite{MinamitsujiDisformalTransformation2020}.
All of them are analytical, although somewhat messy, and have been constructed by performing a disformal transformation on another known (and typically simpler) solution.
Among numerical solutions, we mention the rotating black holes in Einstein--\ae{}ther theory reported in \cite{adam_rotating_2021}, and those in semiclassical gravity of \cite{FernandesRotatingBlack2023}.
Examples that are not technically solutions include the regular black hole models developed in \cite{eichhorn_image_2021,eichhorn_locality-principle_2021} based on a locality principle --- see also \cite{DelaporteParameterizationsBlackhole2022a}.

In all these examples, the inexistence of a Killing horizon is concomitant with the breaking of a little known symmetry called \emph{circularity}.
By definition, a stationary and axisymmetric spacetime is said to be circular if the two Killing vectors are everywhere orthogonal to a family of codimension-two hypersurfaces \cite{CarterBlackHoles1973,CarterRepublicationBlack2009,CarterRepublicationBlack2010}; 
by Frobenius's theorem, this is true if and only if the following conditions are satisfied everywhere in the manifold:
\bse\label{eq:CircCond}
\be
\xi_{[\mu} \psi_\nu \nabla_\rho \xi_{\sigma]} &= 0 \, , \\
\xi_{[\mu} \psi_\nu \nabla_\rho \psi_{\sigma]} &= 0 \, .
\ee
\ese
Though not obvious from the formulation above, these conditions are always satisfied whenever the Ricci tensor is proportional to the metric. 
Hence, in vacuum general relativity, circularity is a consequence of the field equations and can therefore be considered an accidental symmetry.
Beyond general relativity, this is not necessarily the case --- although cf.~\cite{Xie:2021bur} for situations in which it is.
For recent theoretical and phenomenological work on circularity, and the lack thereof, we refer to \cite{AnsonDeformedBlack2021,BabichevTestingDisformal2024a,TakamoriTestingNoncircularity2023,ChenTestingGravity2021,GhoshProbingSpacetime2024,GhoshParameterizedNoncircular2024,DelaporteParameterizationsBlackhole2022a,eichhorn_image_2021,eichhorn_locality-principle_2021,Chen:2023gwm,LongShadowDisformal2020,NakashiRotatingNoncircular2020,ZhouEffectNoncircularity2022}, as well as \cite{BabichevPropertiesGeneral2025}, and references therein.

A series of classical results, summarised by Carter in his excellent lecture notes \cite{CarterBlackHoles1973,CarterRepublicationBlack2009,CarterRepublicationBlack2010}, ensures that if a spacetime is circular then the event horizon is also a Killing horizon.
The proof of this statement is rather geometrical and therefore, though very elegant, it is not particularly transparent.
Recently, however, one of us and a collaborator have been able to solve the differential conditions of \cref{eq:CircCond} and translate them into two algebraic relations among components of the metric 
\cite{BabichevPropertiesGeneral2025}.
Since that result holds within a particular gauge, we choose to renounce, for the most part of this article, to a manifestly coordinate-free notation, and work in that gauge.
Crucially, this will entail no loss of generality.
In so doing, we will be able to show explicitly how \cref{eq:CircCond} being satisfied at least locally on the horizon implies that the horizon is Killing.

\subsection{\label{subsec:gauge}Coordinate choice and horizon characterisation}

We introduce coordinates $v$, $r$, $\theta$, and $\phi$ adapted to the Killing isometries, in the sense that in these coordinates
\be
\xi^\mu \partial_\mu = \partial_v
\qq{and}
\psi^\mu \partial_\mu = \partial_\phi \, ,
\ee
and the components $g_{\mu \nu}(r,\theta)$ of the metric do not depend on $v$ nor $\phi$.
We use general covariance to bring the metric to a \qu{Kerr-like form}\footnote{Note that \qu{Kerr-like} here refers to the Kerr ingoing coordinates, not to the Kerr solution. In particular, the coordinate $v$ is \qu{morally} an advanced time.}, defined by the following conditions:
\be
g_{v\theta} = 0 \, ,
\quad
g_{r\theta} = 0 \, ,
\quad
g_{\theta\phi} = 0 \, ,
\qq{and}
g_{rr} = 0\, .
\ee
That is, we take the line element to read
\be\label{eq:KerrLikeg}
g_{\mu \nu} \dd{x^\mu} \dd{x^\nu} = g_{vv} \dd{v^2} + g_{\theta \theta} \dd{\theta^2} + g_{\phi \phi} \dd{\phi^2}
 + 2 g_{v\phi} \dd{v} \dd{\phi} + 2 g_{vr} \dd{v} \dd{r} + 2 g_{r\phi} \dd{r} \dd{\phi} \, ,
\ee
where the $g_{\mu \nu}$ are arbitrary functions of $r$ and $\theta$.
The fact that \cref{eq:KerrLikeg} entails no loss of generality, i.e.~that this is really a gauge choice and not an Ansatz, is not entirely trivial and has been proven in \cite{BabichevPropertiesGeneral2025}.

We assume that such coordinate system is horizon penetrating, in the sense that the components of the metric are manifestly regular on the horizon --- except perhaps on the axis of symmetry.
Moreover, we assume that the horizon is a connected immersed submanifold specified by the equation $r = H(\theta)$, for some function $H(\theta)$.
We thus introduce
\be\label{eq:zeta}
\zeta_\mu :=& \frac{\alpha}{g^{vr}} \partial_\mu \left( r - H \right) \nonumber\\
=& \frac{\alpha}{g^{vr}} \left( 0, 1, -H', 0 \right)_\mu \, ,
\ee
where $H'$ is the derivative of $H$ with respect to $\theta$ and $\alpha$ is an arbitrary function which we assume to satisfy $\xi^\mu \partial_\mu \alpha = 0 = \psi^\mu \partial_\mu \alpha$.
Such vector is normal to all surfaces $\{r - H = \text{const.}\}$, and for this reason we will often refer to it as \emph{the normal}, for brevity; in particular, it is orthogonal to the horizon.
Note that this vector is naturally defined up to a multiplicative factor: the notation $\alpha/g^{vr}$ reflects this arbitrariness, although we will argue that $\eval{\alpha}_{r=H} = 1$ is perhaps the most natural choice.
Further note that
\be
\zeta_\mu\, \xi^\mu = 0
\qq{and}
\zeta_\mu\, \psi^\mu = 0\, ,
\ee
hence the Killing vectors are tangent to all the hypersurfaces $\{ r-H(\theta) = \text{const.} \}$
For later reference, we also introduce the vector
\be\label{eq:tau}
\tau^\mu := \left( 0, H', 1, 0 \right)^\mu \, ,
\ee
which is such that 
\be
\zeta_\mu\, \tau^\mu = 0
\ee
and therefore it is also tangent to all hypersurfaces $\{ r-H(\theta) = \text{const.} \}$.

Classic results in causality theory ensure that the horizon must be generated by null geodesic segments \cite{HawkingLargeScale1973} --- which, incidentally, may have past endpoints but cannot have future endpoints ---, and therefore, given our assumptions, it must be a null hypersurface.
Hence, the normal must be null on the horizon:
\be\label{eq:normzeta}
\zeta_\mu \, \zeta^\mu = \frac{\alpha^2}{\left( g^{vr} \right)^2 } \left[g^{rr} + \left(H'\right)^2 g^{\theta\theta} \right]
\qq{is such that}
\eval{\zeta_\mu \, \zeta^\mu}_{r=H} = 0\, .
\ee
Clearly, \cref{eq:normzeta} could have multiple roots: the outermost one corresponds to the event horizon, while the others correspond to inner --- generically Cauchy --- horizons; here, our attention is limited to the region exterior to the black hole, extending from infinity up to the outermost horizon, and we mostly disregard the interior --- except for \cref{subsec:tunnel}, in which we will briefly consider the near-horizon region on both sides of the horizon.
Note that, as a consequence of asymptotic flatness, $\zeta_\mu \, \zeta^\mu > 0$ in a neighbourhood of infinity, and in fact everywhere in the domain of outer communication; hence, the hypersurfaces $\{r-H(\theta) = \text{const.}\}$ are timelike outside of the black hole.

It is important to remark that, in general, there exists no linear combination of the Killing vectors that becomes null on the horizon.
Indeed, in the exterior of the black hole, such linear combinations can be found everywhere from infinity up to a surface $r=R(\theta)$ called \emph{rotosurface} and defined by the condition
\be\label{eq:defR}
\eval{\left( \xi_\mu \, \psi^\mu \right)^2 - \left( \xi_\mu \, \xi^\mu \right)\left( \psi_\nu \, \psi^\nu \right)}_{r=R} = 0\, .
\ee
Note that, in our gauge,
\be
\left( \xi_\mu \, \xi^\mu \right)\left( \psi_\nu \, \psi^\nu \right) - \left( \xi_\mu \, \psi^\mu \right)^2 = g_{vv} g_{\phi \phi} - \left( g_{v\phi} \right)^2 =: D_{v\phi} \, ,  
\ee
where we have introduced the symbol $D_{v\phi}$ to denote the determinant of the $v$-$\phi$ sub-block of the metric; moreover
\be
g^{rr} = \frac{g_{\theta\theta} D_{v\phi} }{\det(g_{\mu \nu})} \, .
\ee
Hence, the rotosurface can be equivalently characterised as the locus of points at which $D_{v\phi}$ or $g^{rr}$ vanish.
Since \cref{eq:normzeta} implies $\eval{ g^{rr} }_{r=H} \leq 0$, we deduce that
\be
R(\theta) \geq H(\theta) \, ,
\ee
i.e.~this surface cannot lie inside of the horizon:
generically, it lies outside of it, although the two surfaces may touch at isolated points such as on the axis of symmetry.
Further note that, unless $R=H$, $\{r=R\}$ is a timelike hypersurface.

We point out that the rotosurface has sometimes been called \emph{stationary limit} \cite{AnsonDisformingKerr2021}, since stationary observers can only exist in the region exterior to it.
Indeed, one may understand this surface in analogy with the more familiar ergosurface --- also known as static limit.
For further details on the rotosurface, we refer the reader to \cite{BabichevPropertiesGeneral2025,CarterBlackHoles1973,CarterRepublicationBlack2009,CarterRepublicationBlack2010} and references therein;
here, we merely point out that the phenomenology of rotosurfaces, as separate from horizons and ergosurfaces, appears greatly understudied.  

For later reference, we define 
\be\label{eq:Xi_def}
\Xi^\mu := \xi^\mu + \sigma \, \psi^\mu \, ,
\qq{with}
\sigma := \frac{g^{r\phi}}{g^{vr}} \, .
\ee
Note that
\be\label{eq:sigma_id}
\sigma = \Omega - \frac{g_{r\phi}}{g_{\phi \phi}} \frac{g^{rr}}{g^{vr}} \, ,
\qq{where}
\Omega := - \frac{\xi_\mu \, \psi^\mu}{\psi_\nu \, \psi^\nu} 
\ee
is the angular velocity of frame dragging. 
Though the identity above might not seem trivial, it can be easily verified by writing the components of the inverse metric in terms of the components of the direct metric.
It is worth pointing out that, despite the looks, the vector $\Xi^\mu$ is \emph{not} the linear combination of the Killing vectors that becomes null on the rotosurface, in general, nor is it a Killing vector.
We shall briefly return to this point in \cref{subsec:circiffK}.

The following decomposition is particularly insightful:
\be \label{eq:zeta_decomp}
\zeta^\mu := \alpha \left( 1, \frac{g^{rr}}{g^{vr}}, -H' \frac{g^{\theta\theta}}{g^{vr}}, \frac{g^{r\phi}}{g^{vr}} \right)^\mu = \alpha \left[ \Xi^\mu 
- \frac{H' g^{\theta \theta}}{g^{vr}} \tau^\mu
+  g^{vr} \left( \zeta_\nu\, \zeta^\nu \right) \delta^\mu_{\ r} 
\right] \, ,
\ee
so in particular
\be
\eval{\zeta^\mu}_{r=H} = \eval{\alpha\left[ \Xi^\mu - \frac{H' g^{\theta \theta} }{g^{vr}}  \tau^\mu \right] }_{r=H} \, .
\ee
We point out that, with the choice $\eval{\alpha}_{r=H}=1$, on the horizon $\zeta^\mu$ coincides with $\Xi^\mu$ up to terms $\propto H'$.

Clearly, important simplifications take place if $H' = 0$.
In this case, the horizon is located at a constant value of the radial coordinate $r$.
Moreover, the rotosurface coincides with the horizon, i.e.~$R=H$.
Consequently, the normal $\zeta^\mu$ becomes proportional to the vector $\Xi^\mu$ on the horizon; 
and, since
\be
\eval{\sigma}_{r=H} = \eval{\Omega}_{r=H} 
\quad
\left( H' = 0 \right) \, ,
\ee
this is the only case in which $\Xi^\mu$ is the linear combination of the Killing vectors that becomes null on the rotosurface.
However, we wish to stress that, despite these simplifications, the circularity conditions need not hold and the horizon needs not be Killing.
In the following, we will often refer to this particular case as \emph{minimal} --- in contrast to the generic case $H' \neq 0$, which we will call \emph{not minimal}.

\subsubsection{\label{subsubsec:examples}Examples} 

Though our discussion is general and applies to any stationary and axisymmetric spacetime, for illustrative purposes it will often be useful to refer to specific examples.
Reference \cite{BabichevPropertiesGeneral2025} introduced two examples that are particularly suited to our purposes, as they represent deformations of the Kerr metric in which circularity is broken \qu{softly}.
In order to render the treatment self contained, we recall such examples here.
First of all, we introduce the following notations, relevant for both examples:
\be
\Sigma := r^2 + a^2 \cos^2 \theta \, ,
\quad
\Delta := r^2 - 2M r + a^2 \, ,
\quad
A:= (r^2+a^2)^2 - \Delta a^2 \sin^2 \theta \, ,
\ee
where $M$ and $a$ represent respectively the spacetime's Arnowitt--Deser--Misner (ADM) mass and specific angular momentum.

The first example is \qu{minimal}, in the sense that it falls within the domain of the minimal case introduced above.
The (inverse) metric reads
\be\label{eq:minimal_eg}
g^{\mu \nu} \partial_\mu \partial_\nu &= 
\frac{a^2 \sin^2\theta}{\Sigma} \partial_v \partial_v
+ 2 \frac{a}{\Sigma} \partial_v \partial_\phi
+ \frac{1}{\sin^2 \theta \Sigma} \partial_\phi \partial_\phi
\nonumber\\
&\phantom{=}
\frac{\Delta}{\Sigma} \partial_r \partial_r 
+ \frac{1}{\Sigma} \partial_\theta \partial_\theta
+ 2 \frac{a}{\Sigma} \partial_r \partial_\phi
+ 2 \frac{r^2 + a^2 + \delta}{\Sigma} \partial_v \partial_r \, .
\ee
The function $\delta(r,\theta)$ in the last term parametrises the deviations away from the Kerr solution, with $\delta = 0$ corresponding to said solution.
The spacetime of \cref{eq:minimal_eg} is not circular if $\partial_\theta \delta \neq 0$.
However, the event horizon coincides with the rotosurface: it is located at the largest zero of $\Delta$, i.e.~at $r = r_+$, with $r_+:= M + \sqrt{M^2 - a^2}$, as in the Kerr solution.
Such horizon is not Killing if $\eval{\partial_\theta \delta}_{r=H} \neq 0$.

The second example is \qu{not so minimal}, since it is constructed with the aim of showcasing the difference between the horizon and the rotosurface.
This example therefore falls within the non-minimal case outlined above. 
The (inverse) metric reads
\be\label{eq:nonminimal_eg}
g^{\mu \nu} \partial_\mu \partial_\nu &= 
\frac{a^2 \sin^2\theta}{\Sigma} \partial_v \partial_v
+ 2 \frac{a}{\Sigma} \partial_v \partial_\phi
+ \frac{1}{\sin^2 \theta \Sigma} \partial_\phi \partial_\phi
\nonumber\\
&\phantom{=}
\frac{\tilde{\Delta} }{\Sigma} \partial_r \partial_r 
+ \frac{1}{\Sigma} \partial_\theta \partial_\theta
+ 2 \frac{a}{\Sigma} \partial_r \partial_\phi
+ 2 \frac{r^2 + a^2}{\Sigma} \partial_v \partial_r \, ,
\ee
where
\be
\tilde{\Delta} := r^2 - 2r m(r,\theta) + a^2 
\qq{and}
\tilde{A} := (r^2 + a^2)^2 - \tilde{\Delta} a^2 \sin^2 \theta \, .
\ee
The metric coincides with that of the Kerr solution if $m(r,\theta)$ is a constant.
However, this spacetime is not circular if $\partial_\theta m(r,\theta) \neq 0$.
Reference \cite{BabichevPropertiesGeneral2025} further shows how one can reverse-engineer an $m(r,\theta)$ with the desired asymptotic properties, starting from a generic horizon profile $H(\theta)$.

\subsection{\label{subsec:circiffK}Proof that \qu{circularity \texorpdfstring{$\Leftrightarrow$}{⇔} Killing horizon}}

In the previous subsection, we have characterised the black hole's event horizon, highlighting the fact that --- in general --- it is not a Killing horizon and it does not coincide with the rotosurface.
As mentioned before, these features are connected with the breaking of circularity. 
Here, we therefore show explicitly that if the circularity conditions of \cref{eq:CircCond} are satisfied, at least locally on the rotosurface, then the rotosurface is a null hypersurface and therefore it must coincide with the horizon; 
more specifically, in this case the rotosurface is a constant-$r$ surface.
Moreover, under the same assumption one can further show that the event horizon is also a Killing horizon.
These facts are well known, cf.~\cite{CarterBlackHoles1973,CarterRepublicationBlack2009,CarterRepublicationBlack2010}, but previous proofs were largely geometrical and not particularly transparent; hence, we specify them to our case.

First of all, note that within our gauge the left-hand side of the circularity conditions of \cref{eq:CircCond} can be written as \cite[eqs.~(55)]{BabichevPropertiesGeneral2025}
\bse\be
\label{eq:CircCondGauge1}
C_1 &:= \left[ g_{r\phi} g_{vv} - g_{vr} g_{v \phi} \right] \partial_\theta g^{vr} + \left[ g_{r\phi} g_{v\phi} - g_{vr} g_{\phi \phi} \right] \partial_\theta g^{r\phi} \, , \\
\label{eq:CircCondGauge2}
C_2 &:= \left[ g_{\phi\phi} g_{vv} - \left( g_{v \phi} \right)^2 \right] \partial_\theta g^{vr} + \left[ g_{\phi\phi} g_{v r} - g_{v\phi} g_{r\phi} \right] \partial_\theta g^{rr} \, .
\ee\ese
That is, the spacetime is circular if
\be
C_1 = 0
\qq{and}
C_2 = 0\, .
\ee
Suppose that the circularity conditions are satisfied on the rotosurface $\{ r=R \}$ --- and possibly nowhere else\footnote{The circularity conditions of \cref{eq:CircCond} are always satisfied on the axis of symmetry, since $\psi^\mu$ vanishes there by definition. Here, we mean \qu{nowhere else, except on the axis}.} ---, i.e.
\be
\eval{C_1}_{r=R} = 0
\qq{and}
\eval{C_2}_{r=R} = 0 \, .
\ee
Since, by \cref{eq:defR}, $\eval{ g_{vv} g_{\phi \phi} - (g_{v\phi})^2 }_{r=R} = 0$, \cref{eq:CircCondGauge2} implies one of the following:
\be
\qq*{either}
\eval{ g_{\phi \phi} g_{vr} - g_{v\phi} g_{r\phi} }_{r=R} = 0 \, ,
\qq{or}
\eval{ \partial_\theta g^{rr} }_{r=R} = 0 \, .
\ee
Taking the first of these options, along with $\eval{ g_{vv} g_{\phi \phi} - (g_{v\phi})^2 }_{r=R} = 0$, and recalling that $\eval{g_{vv}}_{r=R} \neq 0$, we deduce
\be
\eval{ g_{v\phi} \left[ g_{v\phi} g_{vr} - g_{vv} g_{r\phi} \right] }_{r=R} = 0 \, .
\ee
This seems to entail that $\eval{C_2}_{r=R} = 0 \Rightarrow \eval{C_1}_{r=R} = 0$, automatically;
however, since
\be
\det(g_{\mu \nu}) = - g_{\theta \theta} \left[ 
g_{r\phi} \left( g_{vv} g_{r\phi} - g_{vr} g_{v \phi} \right) + 
g_{vr} \left( g_{ \phi \phi} g_{vr} - g_{v \phi } g_{r \phi} \right)
\right] \, ,
\ee
the previous equalities also imply that the determinant of the metric vanishes at the rotosurface, i.e.~the metric becomes singular there.
We must therefore discard this first option.
Hence, $\eval{C_2}_{r=R} = 0$ must imply $\eval{ \partial_\theta g^{rr} }_{r=R} = 0$.
In turn, this implies that $R'=0$.
Indeed, by definition, $g^{rr}$ is a constant (namely, zero) over the rotosurface and therefore its derivative in the direction tangent to the rotosurface must be zero.
Such derivative is the following [cf.~\cref{eq:tau}]:
\be
0 = \eval{\left( R' \partial_r + \partial_\theta \right) g^{rr}}_{r=R} \, ;
\ee
hence, as long as $\eval{\partial_r g^{rr}}_{r=R} \neq 0$, we deduce
\be
\eval{ \partial_\theta g^{rr} }_{r=R} = 0 
\Leftrightarrow
R' = 0 \, .
\ee
But if $R'=0$ then the rotosurface is a null hypersurface.
Indeed, the norm of the vector normal to the rotosurface [cf.~\cref{eq:normzeta}] vanishes
\be
\eval{g^{rr} + \left( R' \right)^2 g^{\theta \theta} }_{r=R} = 0\, .
\ee
Therefore, if the circularity conditions are satisfied on $r=R$, the rotosurface must coincide with the horizon, i.e.~$R=H$, and be a constant-$r$ surface, i.e.~$R' = H' = 0$.

Under the assumptions above, it is relatively simple to prove that the angular velocity of frame dragging $\Omega$ is a constant over the horizon --- meaning that the derivatives in the directions tangent to the horizon all vanish.
This result is truly standard, and a proof can be found in e.g.~\cite[sec.~6.3.4]{FrolovBlackHole1998}, which is the same as \cite[sec.~4]{CarterBlackHoles1973,CarterRepublicationBlack2010}.\footnote{
We advise the reader that the statement reported in these references should be taken with some grain of salt. What is proven there is that $\eval{\nabla_\mu \Omega}_{r=H}$ lies in the span of $\xi_\mu$ and $\psi_\mu$; ref.s \cite{FrolovBlackHole1998,CarterBlackHoles1973,CarterRepublicationBlack2010} then go on to claim that this implies $\eval{\nabla_\mu \Omega}_{r=H} = 0$, which is --- we believe --- not correct. Though it is true that $\eval{\xi^\mu \nabla_\mu \Omega}_{r=H}=\eval{\psi^\mu \nabla_\mu \Omega}_{r=H}=0$, this merely implies that $\eval{\nabla_\mu \Omega}_{r=H}$ is a null vector --- not that it vanishes. In other words, the gradient of $\Omega$ is normal to the horizon. A function whose gradient truly vanishes is $\Omega_H$ as we define it below.}
The key consequence of this fact is that one can find a Killing vector which becomes null on the horizon, is orthogonal to it and therefore is a generator.
The event horizon is thus also a Killing horizon.
Such Killing vector is $\xi^\mu + \Omega_H \psi^\mu$, where we have called
\be
\Omega_H := \eval{ \Omega }_{r=H} \, ;
\ee
more precisely, $\Omega_H$ is a function that coincides with $\Omega$ on the horizon and is constant in the direction orthogonal to it.
It is worth pointing out that, since [cf.~\cref{eq:sigma_id}]
\be
\eval{\sigma}_{r=H} = \eval{ \Omega }_{r=H}
\quad 
\left( H' = 0 \right) \, ,
\ee
the vector $\Xi^\mu$ of \cref{eq:Xi_def} coincides with the Killing vector on the horizon, and yet it is \emph{not} itself Killing.

\section{\label{sec:surfgrav}Surface gravities}

The notion of surface gravity is key in a thermodynamic interpretation of black holes, due to its identification with the horizon's temperature.
Actually, the term \qu{surface gravity} may refer to a swath of technical definitions embodying logically distinct notions \cite{CroppSurfaceGravities2013}.
In the simplest situations --- e.g.~for the Kerr solution ---, all these notions coincide, and the various definitions are but equivalent ways of computing the same quantity.
However, in less simple settings, differences can become evident.

In this section we shall thus compute expressions for a selection of popular \qu{surface gravities}, and describe their relations.
We start by examining two closely related notions: the \emph{normal} and the \emph{inaffinity} surface gravity, both of which refer to the null geodesics that generate the horizon.
We then move on to the so-called \emph{peeling} surface gravity, which instead captures the behaviour of outgoing causal geodesics in the close vicinity of the horizon.

\subsection{\label{subsec:knorm}On-horizon surface gravities}

As mentioned, a classical result in causality theory \cite{HawkingLargeScale1973} states that the horizon is generated by null geodesic segments.
Hence, at least on each connected component of the horizon, it mus be possible to define a quantity $\ki$ such that
\be\label{eq:ki}
\eval{\zeta^\nu \nabla_\nu \zeta_\mu }_{r=H} =: \ki \, \eval{ \zeta_\mu }_{r=H} \, .
\ee
We will refer to this quantity, which measures the amount by which the horizon's generators fail to be affinely parametrised, as \emph{inaffinity} surface gravity.

On the other hand, the fact that the normal is null only on the horizon, and nowhere else, means that the equation $\eval{\zeta_\mu \, \zeta^\mu }_{r=H} = 0$ may be used to specify the horizon's location, and it thus plays the very same role as the equation $r - H = 0$.
Hence, the gradient $\nabla_\mu \left( \zeta_\nu \, \zeta^\nu \right)$ must be normal to the horizon and proportional to $\zeta_\mu$ on the horizon.
This entails that there must exist a quantity $\kn$ such that
\be\label{eq:kn}
\eval{ \nabla_\mu \left( \zeta_\nu \, \zeta^\nu \right) }_{r=H} =: 2 \kn \, \eval{ \zeta^\mu }_{r=H} \, .
\ee
We will refer to $\kn$ as \emph{normal} surface gravity.

Both definitions rely on properties of null hypersurfaces and of their normal vectors, and they are technically consistent;
what is not entirely obvious at this point is whether any of them is relevant --- i.e.~whether they play the role that the surface gravity usually plays.
For instance, one way in which our definitions differ from the familiar ones, given in terms of the Killing vector generating the Killing horizon, is that such Killing vector is timelike outside of the horizon --- while our $\zeta^\mu$ is spacelike.
Hence, one might suspect that $\ki$ and $\kn$ as defined in \cref{eq:ki,eq:kn} might not represent physically interesting quantities.
Further note that \cite{BabichevPropertiesGeneral2025} already computed an inaffinity surface gravity, making use of the equivalent of our vector $\Xi^\mu$, and it is not obvious that our definition agrees with that one.
In the following, will we show that our definitions reduce to known results when they should, and argue that they provide reasonable extensions of the usual notions beyond familiar cases.

What is evident already at this point, instead, is that the two notions introduced here are logically distinct and, \textit{a priori}, independent.
However, a few lines of algebra allow to establish an insightful relation that connects them.
Such relation hinges on the fact that the horizon is an immersed submanifold and, consequently, that the normal is a hypersurface-orthogonal vector; it might therefore be violated in topologically non-trivial configurations, but it does hold for the kind of horizons analysed in this article.

To derive said relation, recall that 
\be
\zeta_\mu = \frac{\alpha}{g^{vr}} \partial_\mu \left( r - H \right) \, .
\ee
We have
\be
\nabla_\mu \zeta_\nu &= \frac{\alpha}{g^{vr}} \nabla_\mu \partial_\nu \left( r - H \right) + \zeta_\nu \partial_\mu \log(\frac{\alpha}{g^{vr}}) \, ;
\ee
the derivatives in the first term commute, since $r-H$ is a scalar function, hence\footnote{Note that this expression is precisely the statement that the twist of $\zeta^\mu$ vanishes, as it is a hypersurface-orthogonal vector field --- in accordance with Frobenius' theorem \cite{Wald_GR}.}
\be\label{eq:anti_nabla_zeta}
\nabla_{[\mu } \zeta_{\nu ]} = - \zeta_{[\mu } \partial_{\nu]} \log(\frac{\alpha}{g^{vr}}) \, .
\ee
Unpacking the derivative in the definition of $\kn$, we find
\be
\nabla_\mu \left( \zeta_\nu \, \zeta^\nu \right) &= 2 \zeta^\nu \nabla_\mu \zeta_\nu \nonumber\\
&= 2 \zeta^\nu \nabla_\nu \zeta_\mu  - 4 \zeta^\nu \zeta_{[ \mu } \partial_{\nu ]} \log(\frac{\alpha}{g^{vr}}) \nonumber\\
\Rightarrow 
\eval{\nabla_\mu \left( \zeta_\nu \, \zeta^\nu \right)}_{r=H} &= \eval{ 2 \zeta^\nu \nabla_\nu \zeta_\mu  - 2 \zeta_{\mu } \zeta^\nu \partial_{\nu} \log(\frac{\alpha}{g^{vr}}) }_{r=H}
\, .
\ee
We thus arrive to the following:
\be\label{eq:kikn_rel}
\ki = \kn + \eval{ \zeta^\nu \partial_{\nu } \log(\frac{\alpha}{g^{vr}}) }_{r=H} \, .
\ee
That is, the mismatch between these two notions of surface gravity is controlled by the derivative of $\alpha/g^{vr}$ in the direction normal to the horizon.
Incidentally, this entails that, in principle, one can always choose $\alpha$ so that $\kn=\ki$: $\alpha = g^{vr}$ trivially accomplishes this task.
However, if one has a physical argument for picking a less trivial normalisation for $\zeta^\mu$, one might find that $\kn$ and $\ki$ differ.
In \cref{subsec:gauge}, we already hinted to the fact that $\eval{\alpha}_{r=H}=1$ appears to be the most natural choice; we anticipate that for this choice, in general, we will find $\ki \neq \kn$.

Given the Kerr-like gauge of \cref{eq:KerrLikeg}, we are now in the position to compute the explicit expressions of $\ki$ and $\kn$ in terms of the components of the metric.
Starting from $\kn$, we get
\be
\nabla_\mu \left( \zeta_\nu \, \zeta^\nu \right) &= \left(\frac{\alpha}{g^{vr}} \right)^2  \left( 0, \partial_r g^{rr} + \left( H' \right)^2 \partial_r g^{\theta \theta}, \partial_\theta g^{rr} + \left( H' \right)^2 \partial_\theta g^{\theta \theta} + 2 H' H'' g^{\theta \theta} , 0 \right)_\mu  \nonumber\\
&\phantom{=} + \left( \zeta_\nu \, \zeta^\nu \right) \partial_\mu 2 \log(\frac{\alpha}{ g^{vr} }) \, .
\ee
The second term vanishes when evaluated on the horizon; to check that the first term is proportional to $\zeta_\mu$ on the horizon, we recall that [cf.~\cref{eq:tau}]
\be
0 &= \eval{ \tau^\mu \partial_\mu \left( \zeta_\nu \, \zeta^\nu \right) }_{r=H} \nonumber\\
&= \eval{ H' \left[\partial_r g^{rr} + \left( H' \right)^2 \partial_r g^{\theta \theta} \right] + \left[ \partial_\theta g^{rr} + \left( H' \right)^2 \partial_\theta g^{\theta \theta} + 2 H' H'' g^{\theta \theta} \right] }_{r=H} \, .
\ee
We can thus read off
\be\label{eq:kn_comp_alpha}
\kn = \eval{ \frac{\alpha}{2g^{vr}} \left[\partial_r g^{rr} + \left( H' \right)^2 \partial_r g^{\theta \theta} \right] }_{r=H} \, .
\ee
On the other hand, since we assumed $\xi^\mu \partial_\mu \alpha = 0 = \psi^\mu \partial_\mu \alpha$,
\be
\zeta^\mu \partial_\mu \log( \frac{\alpha}{g^{vr}} ) =  g^{rr} \partial_r \left( \frac{\alpha}{g^{vr}} \right) - H' g^{\theta \theta} \partial_\theta \left( \frac{\alpha}{g^{vr}} \right)
\, .
\ee
Hence
\be\label{eq:ki_comp_alpha}
\ki = \eval{ 
\frac{\alpha}{2g^{vr}} \left[ \partial_r g^{rr} + \left( H' \right)^2 \partial_r g^{\theta \theta} \right]
+ g^{rr} \partial_r \left( \frac{\alpha}{g^{vr}} \right) - H' g^{\theta \theta} \partial_\theta \left( \frac{\alpha}{g^{vr}} \right)
}_{r=H} \, .
\ee

A rapid inspection of \cref{eq:kn_comp_alpha,eq:ki_comp_alpha} shows that the explicit values of $\ki$ and $\kn$, as well as that of their difference, depend on the choice of the normalisation $\alpha$.
This ambiguity has nothing to do with circularity, or the lack thereof, since it exists also in, say, the Kerr solution: there too the value of the surface gravity may be rescaled at will by simply rescaling the Killing vector used to define it.
What picks the \qu{right} value in the Kerr case is the normalisation of the Killing vector \emph{at infinity}, $\xi_\mu\, \xi^\mu \to -1$ as $r \to \infty$. 
Classically, such ambiguity appears rather inconsequential; quantum mechanically, the choice of the Killing time at infinity selects the vacuum state, hence providing a physical argument for fixing $\alpha$ might be relevant.

The choice we hinted to in \cref{subsec:gauge}, i.e.~$\eval{\alpha}_{r=H} = 1$, is ultimately justified by the decomposition of \cref{eq:zeta_decomp}.
In particular, this choice entails that, in the case $H'=0$, $\zeta_\mu$ coincides with the vector $\Xi_\mu$ on the horizon, and therefore it seems to provide the most natural generalisation of known cases.

Indeed, in the minimal case one could very reasonably define the surface gravities in terms of $\Xi^\mu$, instead of $\zeta^\mu$.
Incidentally, this is the strategy adopted in \cite{BabichevPropertiesGeneral2025} to compute the inaffinity surface gravity of the minimal example of \cref{subsubsec:examples}.
Specifically, one could define $\ki^\Xi$ and $\kn^\Xi$ such that
\be
\eval{ \Xi^\mu \nabla_\mu \Xi_\nu }_{r=H} = \eval{ \ki^\Xi \Xi_\nu }_{r=H} 
\qq{and}
\eval{ \nabla_\mu \left( \Xi_\nu\, \Xi^\nu \right) }_{r=H} = \eval{ - 2 \kn^\Xi \Xi_\mu }_{r=H}\, . 
\ee
Since, in the minimal case, one has
\be
\zeta^\mu = \alpha \left[ \Xi^\mu + g^{vr} \left( \zeta_\nu \, \zeta^\nu \right) \delta^\mu_{\ r} \right]
\quad
\left( H'=0 \right) \, ,
\ee
a few lines of algebra suffice to show that
\be
\eval{ \zeta^\mu \nabla_\mu \zeta_\nu }_{r=H} = \eval{ \alpha^2 \Xi^\mu \nabla_\mu \Xi_\nu }_{r=H}
\qq{and}
\eval{ - \nabla_\mu \left( \zeta_\nu\, \zeta^\nu \right) }_{r=H} = \eval{ \alpha^2 \nabla_\mu \left( \Xi_\nu\, \Xi^\nu \right) }_{r=H} \, ;
\ee
this gives 
\be
\ki = \eval{\alpha^2}_{r=H} \ki^\Xi
\qq{and}
\kn = \eval{\alpha^2}_{r=H} \kn^\Xi\, .
\ee
Hence, $\eval{\alpha}_{r=H}=1$ ensures that our definitions agree with more familiar situations.

Specifying to $\eval{\alpha}_{r=H} = 1 $, we get
\be
\label{eq:kn_comp}
\kn &= \eval{  \frac{\partial_r g^{rr} + \left( H' \right)^2 \partial_r g^{\theta \theta}}{2 g^{vr}} }_{r=H} \, ,\\
\label{eq:ki_comp}
\ki &= \eval{ 
\frac{\partial_r g^{rr} + \left( H' \right)^2 \partial_r g^{\theta \theta}}{2 g^{vr}}
- \frac{g^{rr} \partial_r g^{vr} - H' g^{\theta\theta} \partial_\theta g^{vr}}{\left( g^{vr} \right)^2}
}_{r=H} \, .
\ee
Further note that, thanks to the decomposition of \cref{eq:zeta_decomp} and recalling that $\xi^\mu\partial_\mu g^{vr} = 0 =\psi^\mu\partial_\mu g^{vr}$, we may write
\be \label{eq:kikn_rel_comp}
\ki = \kn + \frac{H' g^{\theta\theta}}{\left( g^{vr} \right)^3} \tau^\mu \partial_\mu g^{vr} \, .
\ee
An important remark is the following: in the case $H' = 0$, we have that
\be \label{eq:kappa_minimal}
\ki = \kn
\quad
\left( H'=0 \right)\, .
\ee
This can be seen from \cref{eq:kn_comp,eq:ki_comp} by setting $H' = 0$ and recalling that in this case $g^{rr}$ vanishes at the horizon; 
and even more directly from \cref{eq:kikn_rel_comp}.
We stress that an horizon with $H' = 0$ can still be not Killing, hence the fact that $\ki$ and $\kn$ coincide is not at all trivial.

To close this section, we report the values of $\ki$ and $\kn$ computed for the examples of \cref{subsubsec:examples}.
For the \qu{minimal} example, we find
\be \label{eq:minimal_kappa_eg}
\ki = \kn = \frac{r_+ - M}{r_+^2 + a^2 + \delta} \, ,
\ee
which agrees with \cite[eq.~(5.19)]{BabichevPropertiesGeneral2025};
note that this quantity depends on the angle $\theta$ if $\delta$ does.
For the \qu{not-so-minimal} case, we find
\be \label{eq:nonminimal_kappa_eg}
\kn &= \eval{ \frac{r - m - r \partial_r m}{r^2 + a^2} - \frac{r \left( H' \right)^2 }{\Sigma \left( r^2 + a^2 \right)} }_{r=H} \, ,\\
\ki &= \kn + \eval{H' \frac{2 a^2 \sin\theta \left[ \left( r^2 + a^2 \right) - H' r \sin\theta \right] }{ \left( r^2 + a^2 \right)^3}}_{r=H} \, ;
\ee
note that the two notions disagree, as expected.

\subsection{\label{subsec:kpeel}Peeling surface gravity}

A markedly different notion of surface gravity relates to the behaviour of outgoing future-directed causal geodesics in the vicinity of the horizon.
As we will show, such geodesics \qu{peel off} of the horizon, in a suitable sense, and for this reason one usually speaks of \emph{peeling surface gravity}.
Notably, this notion seems to be the most relevant one for what concerns Hawking radiation \cite{BarceloMinimalConditions2011,BarceloHawkinglikeRadiation2011}.

To set up our analysis, we consider geodesics passing through a generic point $P$, located outside of the horizon, then take the limit in which $P$ approaches the horizon.
We adopt the language of Hamiltonian dynamics, i.e.~we regard geodesics as the trajectories of point particles with mass $\mu$, position $x^\mu$, conjugate momentum $k_\mu$, and subject to the Hamiltonian
\be\label{eq:Hamiltonian}
H \left( x^\mu ,\, k_\mu \right) = \frac{1}{2} g^{\mu \nu} \left( x^\alpha \right) k_\mu k_\nu \, .
\ee
Here, $x^\mu$ and $k_\mu$ are understood as functions of a parameter that varies monotonically along the trajectory and whose choice is essentially arbitrary.
For our purposes, it is particularly convenient to parametrise the trajectory in terms of
\be
z:= r - H(\theta) \, ,
\ee
so that the limit in which $P$ approaches the horizon corresponds to $z \to 0$.
(We should point out that $z$ might not be a good parameter for \emph{all} trajectories through $P$, but it is so for the ones we focus on in this section.)
A derivative with respect to $z$ will be denoted by an overdot, so that Hamilton's equations read
\bse\label{eq:xdotkdot}
\be
\label{eq:xdot}
\dot{x}^\mu &= g^{\mu \nu} k_\nu \, , \\
\label{eq:kdot}
\dot{k}_\mu &= - \frac{1}{2} \partial_\mu g^{\alpha \beta} k_\alpha k_\beta \, .
\ee\ese
Due to the Killing symmetries, the Killing energy and angular momentum
\be\label{eq:KillingEL}
-\omega := k_\mu \, \xi^\mu 
\qq{and}
m := k_\mu \, \psi^\mu 
\ee
are conserved along the trajectory;
the norm of the momentum, i.e.~the particle's mass,
\be\label{eq:mass}
- \mu^2 = g^{\mu \nu} k_\mu k_\nu
\ee
is also conserved.
Thus, in general, there exist three independent integrals of motion.

In specific situations, a fourth integral might appear.
If this is the case, the system is completely integrable in the sense of Liouville --- i.e.~it has as many integrals of motion as degrees of freedom in configuration space.
As a consequence, the Hamilton--Jacobi equation corresponding to \cref{eq:Hamiltonian} admits separation of variables, and the trajectories can be solved \qu{by quadrature}, i.e.~with no need to integrate differential equations.
Clearly, therefore, complete integrability entails a substantial simplification of the problem; however, it is a rather rare property:
though several metrics of interest, including the Kerr solution, do enjoy it, complete integrability seems to require the circularity of the spacetime, and for this reason we cannot assume it here --- see e.g.~\cite{BenentiRemarksCertain1979,JohannsenRegularBlack2013,DelaporteParameterizationsBlackhole2022a,BabichevPropertiesGeneral2025} and references therein.

Nonetheless, the high degree of symmetry exhibited by the system is sufficient for our purposes.
Introducing the notations
\be\label{eq:betagamma}
\beta :=  g^{vr} \omega - g^{r\phi} m
\qq{and}
\gamma^2 := - \frac{ g_{\theta \theta} \left( g_{r\phi} \omega + g_{vr} m \right)^2 }{\det(g_{\mu\nu})} \, ,
\ee
the (conserved) normalisation of the momentum, \cref{eq:mass}, can be spelled out as
\be\label{eq:mass_shell}
g^{rr} \left( k_r \right)^2 - 2 \beta k_r + g^{\theta \theta} \left( k_\theta \right)^2 + \left( \mu^2 + \gamma^2 \right) = 0 \, .
\ee
This equation allows to express one component of the momentum, e.g.~$k_\theta$, in terms of the particle's position, of the conserved quantities, and of the remaining component of the momentum, say $k_r$.
Hence, thanks to \cref{eq:mass_shell}, in order to determine the momentum completely one only needs to solve one component of \cref{eq:kdot}, which with the notations introduced above reads
\be\label{eq:kdot_comp}
\dot{k}_\mu = -\frac{1}{2} \left[ \left( k_r \right)^2 \partial_\mu g^{rr} - 2 k_r \partial_\mu \beta + \left( k_\theta \right)^2 \partial_\mu g^{\theta \theta} + \partial_\mu \gamma^2
\right] \, .
\ee

Since the horizon is a causal boundary, we expect that the solutions to Hamilton's \cref{eq:xdotkdot} that represent outwards-moving trajectories should not be analytic at the horizon --- since, if they were, there would exist future-directed causal geodesics that exit the horizon, which is a contradiction.
For such trajectories, we expect that at least one of the components of the momentum should diverge in the limit in which the point $P$ approaches the horizon, i.e.~as $z \to 0$.
When focusing on such case, \cref{eq:mass_shell,eq:kdot_comp} should therefore be understood as asymptotic relations.

Using the following notations
\be \label{eq:K_b_Gamma}
K:= \partial_r g^{rr} - \frac{g^{rr}}{g^{\theta\theta}} \partial_r g^{\theta\theta} \, ,
\quad
b:= \partial_r \beta - \frac{\beta}{g^{\theta \theta}} \partial_r g^{\theta\theta} \, ,
\quad
\Gamma := \partial_r \gamma^2 - \frac{\gamma^2 + \mu^2}{g^{\theta \theta}} \, ,
\ee
Hamilton's equation for $k_r$ can be written as 
\be
\dot{k}_r = - \frac{1}{2} \left[ K \left( k_r \right)^2 - 2b \, k_r + \Gamma \right] \, .
\ee
Its solution has the structure
\be
k_r = \frac{2}{K_H z} + \left( \frac{b_H}{K_H} - \frac{\dot{K}_H}{K_H^2} \right) + \sum_{n=1}^\infty \rho_n z^n \, ,
\ee
where the coefficients $\rho_n$ can be computed order by order in $z$. 
The suffix $H$ indicates that the relative quantity is evaluated on the horizon.
In particular, $k_r$ displays a simple pole whose residue is controlled by the value of $K$ evaluated on the horizon; the next-to-leading term is finite and not zero, while the remainder vanishes polynomially with $z$. 

The behaviour of $k_\theta$ can be read off via the normalisation of the momentum:
\be \label{eq:k_theta}
\left( k_\theta \right)^2 &= 
4 \frac{\left( H' \right)^2}{K_H^2} \frac{1}{z^2}
+ \eval{\frac{4K \left( \beta K - \dot{g}^{rr} \right) - 4 \left( H' \right)^2 \left[ K \dot{g}^{\theta\theta } + g^{\theta \theta}\left( - b K + \dot{K} \right)\right]}{g^{\theta\theta} K^3}}_{r=H} \frac{1}{z} \nonumber\\
&\phantom{=} + \order{1} \, .
\ee
Notably, the leading divergence of $k_\theta$ depends quite crucially on whether $H'=0$ or not.
Indeed, if $H' \neq 0$, we simply have [the sign can be easily determined from \cref{eq:kdot}]
\be
k_\theta = - 2 \frac{H'}{K_H z} + \order{1}\, .
\ee
If $H'=0$, we notice that the coefficients in front of both the $1/z^2$ and $1/z$ terms in \cref{eq:k_theta} vanish for $z \to 0$. In fact, in this case, $z=r- H$ is just a constant shift in $r$. From Hamilton's equations we can write
\be
1= \frac{{\rm d} r}{{\rm d} z}= \dot r = g^{rr} k_r - \beta = \eval{ \left[ \frac{2\partial_r g^{rr}}{K}- \beta \right]}_{r=H} + \order{z} \,.
\ee
The condition $H'=0$ also implies $K_H= \eval{\partial_r g^{rr} }_{r=H}$, thus
\be \label{eq:prefactor}
\beta_H=1 \Rightarrow \beta_H K_H - \dot{g}^{rr}|_{r=H}= K_H- \partial_r g^{rr}|_{r=H}=0 \,.
\ee
Therefore, no power-law divergences are present in $k_\theta$ for $z \to 0$. Its behaviour can be derived by integrating  explicitly the relative Hamilton's equation near $z=0$ upon setting $H'=0$:
\be \label{eq:k_theta_minimal}
\dot{k}_\theta = -\frac{1}{2} \partial_\theta g^{rr} k_r^2 + \partial_\theta \beta k_r + \order{1} 
\Rightarrow k_\theta 
=\eval{\frac{K \partial_\theta \beta- \partial_\theta K }{K^2}}_{r=H} \log(z)
+ \order{1}\, .
\ee
Coherently with what found in \cref{eq:prefactor}, $k_\theta$ shows no power-law divergence at the horizon; instead, we find a logarithmic divergence. Two remarks are in order. First of all, one can check that in the (circular) case of the Kerr spacetime, this logarithmic behaviour disappears, since the $\theta$-dependence factorizes
\be
(K \partial_\theta \beta- \partial_\theta K)_{r=H}\overset{\rm Kerr}{=}- \eval{\frac{\partial_\theta \Sigma}{\Sigma^2}}_{r=H}(K_H \beta_H-K_H)=0.
\ee
This correctly shows the finiteness of $k_\theta$ at the Killing horizon and serves as a consistency check of \cref{eq:k_theta_minimal} for the minimal breaking case. Secondly, we point out that such logarithmic behaviour in $k_\theta$ is completely subleading with respect to the divergence of $k_r$. As a consequence, we shall see in \cref{subsec:tunnel} that the presence of the logarithm in \eqref{eq:k_theta_minimal} will not affect the radiative properties of the black hole. 

It is rather straightforward now to compute how the coordinates diverge along the trajectory.
In particular, we have:
\be
\dot{v} &= g^{vr} k_r - \omega g^{vv} + m g^{v\phi} = 2 \eval{\frac{g^{vr}}{K} }_{r=H} \frac{1}{z} + \order{z^0}\\
\Rightarrow v &= 2 \eval{\frac{g^{vr}}{K} }_{r=H} \log(z) + \order{z} \, .
\ee
So, we naturally define the \emph{peeling} surface gravity as 
\be \label{eq:k_peeling}
\frac{1}{\kp} := 2 \eval{\frac{g^{vr}}{K} }_{r=H} \, .
\ee
Recalling the definition of $K$, \cref{eq:K_b_Gamma}, we have
\be
\kp = \eval{  \frac{\partial_r g^{rr} + \left( H' \right)^2 \partial_r g^{\theta \theta}}{2 g^{vr}} }_{r=H}\, ,
\ee
which is exactly the expression we found when discussing the normal surface gravity in \cref{eq:kn_comp} --- provided we normalise the normal vector with $\eval{\alpha}_{r=H} = 1$.
We should point out, however, that the definition of $\kp$ is also ambiguous, in the same way as that of $\kn$ is: should one rescale the Killing coordinate $v$, one would still find a logarithmic peeling off of the horizon, but the value of $\kp$ would similarly be rescaled.

Hence, the upshot of this discussion is that, provided normalisations are chosen coherently, one can make the general claim
\be
\kp = \kn \, .
\ee
This coincidence is quite striking and not at all expected.

\section{\label{sec:thermo}Towards thermodynamics}

Having developed the three notions of surface gravity, and established the mutual relations among them, we are in the position of taking some further steps towards thermodynamics.
We shall start by deriving Smarr's formula, which, being a relation among mass, angular momentum, and horizon's area, constitutes the technical underpinning for the first law.
In particular, we will comment on the difficulties brought about by the fact that the generators of the horizons herein considered are generically not Killing vectors. 
Despite these complications, we will then perform a simple computation showing that these horizons do emit Hawking radiation at a \qu{temperature} set by the value of the peeling surface gravity.
Finally, we will collect thoughts and remarks on the four laws of black hole thermodynamics, highlighting the difficulties inherent to the non-Killing case.

\subsection{\label{subsec:Smarr}Smarr's formula}

An intermediate technical step towards black hole thermodynamics is the derivation of Smarr's formula \cite{SmarrMassFormula1973,BardeenFourLaws1973}. 
As a relation among mass, angular momentum and horizon's area, this formula represents the main building block of the so-called first law --- namely, the law expressing energy conservation.
Here, we present a derivation adapted to the case in which the horizon is not generated by a Killing vector. 
We will then return to the relationship between such formula and the first law in \cref{subsec:4law}.

We start by considering the following quantity:
\be \label{eq:I}
I := \int_{\mathcal S} \zeta_{[\mu } N_{\nu]} \nabla^\mu \zeta^\nu \, .
\ee
Here, $\mathcal S$ is a bidimensional spacelike section of the horizon;
the integration measure $ \sqrt{h} \dd[2]{x} $, with $h_{\mu \nu}$ the induced metric on $\mathcal S$, is implied;
the vector $N^\mu$ is normal to the section $\mathcal{S}$, and we choose it to satisfy
\be \label{eq:N}
N_\mu \, N^\mu
\qq{and}
N_\mu \, \zeta^\mu = -1 \, ;
\ee
note that $\zeta^\mu$ and $N^\mu$ are both null and normal to $\mathcal{S}$, but they are linearly independent.

The derivation of Smarr's formula consists in evaluating \cref{eq:I} in two different ways. 
On the one hand, we have
\be \label{eq:LHS}
I = \frac{1}{2} \int_{\mathcal S}  \zeta_{\mu } N_{\nu} \nabla^\mu \zeta^\nu - \zeta_{\nu } N_{\mu} \nabla^\mu \zeta^\nu= \int_{\mathcal S} \frac{\kn-\ki}{2} \, .
\ee
On the other hand, decomposing $\zeta^\mu$ as in \cref{eq:zeta_decomp}, we get
\be \label{eq:RHS}
I = \int_{\mathcal S} \zeta_{[\mu } N_{\nu]} \nabla^\mu  \left[ \alpha \left( 
\Xi^\nu 
- \frac{H' g^{\theta \theta}}{g^{vr}} \tau ^\nu 
+ \frac{ g^{vr} }{\alpha^2}  \left( \zeta_\rho \, \zeta^\rho \right) \delta^\nu_r 
\right) \right]\, .
\ee
The rest of the derivation will consist in unpacking these expressions and assigning a physical meaning to the various terms. 

Before moving on, however, a remark is in order.
At first sight, both \cref{eq:LHS} and \cref{eq:RHS} seem to depend quite drastically on the choice $\alpha$: specifically, not only on the value of $\alpha$ at the horizon, but also on that of its gradient --- i.e.~on $\alpha$ away from the horizon.
A quick computation shows that this conclusion is premature.
Indeed, confronting with \cref{subsec:knorm}, we recall
\be \label{eq:ki-kn}
\frac{\kn - \ki}{2} = - \frac{1}{2} \zeta^\mu \partial_\mu \log \left( \frac{\alpha}{g^{vr}} \right) \, ,
\ee
hence \cref{eq:LHS} amounts to the integral of the expression above.
On the other hand, focusing on \cref{eq:RHS}, when the derivative acts on $\alpha$ we get
\be \label{eq:d_alpha}
I \supset \int_{\mathcal S} \zeta_{[\mu } N_{\nu]} (\nabla^\mu \alpha)  \frac{\zeta^\nu}{\alpha}=-\frac 1 2 \int_{\mathcal S} \zeta^{\mu } \partial_\mu \log(\alpha) \, .
\ee
Equating the two, we realise that the gradient of $\alpha$ cancels, and only the value $\eval{\alpha}_{r=H}$ bears some impact on the interpretation of the formula.
Since we have argued that setting $\eval{\alpha}_{r=H} = 1$ appears rather natural, here we decide to extend this choice away from the horizon and take $\alpha \equiv 1$ --- i.e., in particular, $\eval{\nabla_\mu \alpha}_{r=H}=0$.
According to this argument, this extension entails no loss of generality with respect to what we have already assumed before.

Setting $\alpha \equiv 1$ from here on, we can proceed with our derivation by unpacking the various terms of \cref{eq:RHS}. 
The first one is
\be \label{eq:RHS_1}
 \int_{\mathcal S} \zeta_{[\mu } N_{\nu]} \nabla^\mu \Xi^\nu
=  \int_{\mathcal S} \zeta_{[\mu } N_{\nu]} \left[ \nabla^\mu \xi^\nu + \sigma \nabla^\mu \psi^\nu  + \psi^\nu \nabla^\mu \sigma \right]\, .
\ee
As customary, we define the Komar mass $M_H$ and the Komar angular momentum $J_H$ as
\be
M_H := -\frac{1}{4 \pi G} \int_{\mathcal S} \zeta_{[\mu } N_{\nu]} \nabla^\mu \xi^\nu 
\qq{and} 
J_H: =\frac{1}{8 \pi G} \int_{\mathcal S} \zeta_{[\mu } N_{\nu]} \nabla^\mu \psi^\nu \,;
\ee
moreover, we define 
\be
\overline{\sigma}_H := \frac{1}{8 \pi J_H} \int_{\mathcal S} \zeta_{[\mu } N_{\nu]} \sigma_H \nabla^\mu \psi^\nu \, .
\ee
(We have introduced the notation $\sigma_H:= \eval{\sigma}_{r=H}$ to stress that the integrand is evaluated on the horizon, and for consistency with \cref{subsec:tunnel}.)

We emphasize that both $M_H$ and $J_H$ are conserved quantities, as it happens in the circular case. 
This is because both $\xi^\mu$ and $\psi^\mu$ are Killing vectors, generating respectively time translations and rotations around the axis, and the proof of their conservation holds for non-circular spacetimes without modifications --- see e.g.~\cite{Poisson:2009pwt}. 
We also remark that the mass $M_H$ represents the total energy of the spacetime associated with $\xi^\mu$; this term may take into account non-gravitational sources of energy, as it happens for an electrically charged black hole. 
As we explained before, non-circular black holes are not vacuum solutions of general relativity, and they might be characterized by different macroscopic charges. In principle, these can be contained in $M_H$.

The second term of \eqref{eq:RHS} is genuinely new, in the sense that it does not arise if the horizon is Killing, and in principle it does not vanish.
The last term instead gives 
\be
\int_{\mathcal S} \zeta_{[\mu } N_{\nu]} g^{vr} \delta^\nu_r \nabla^\mu |\zeta|^2=   \int_{\mathcal S} \zeta_{r } g^{vr}  \kn = \int_{\mathcal S}     \kn\, .
\ee

Putting everything together, we can finally write
\be \label{eq:Smarr_general}
 \overline{\kappa} \mathcal{A}_H=  4 \pi G M_H - 8 \pi G \, \overline{\sigma}_H J_H + \eta\,,
\ee
where we have introduced the notations
\be
\label{eq:kmean}
\overline{\kappa} &:= \frac{1}{  \mathcal{A}_H } \int_{\mathcal S} \kn \, ,
\ee
and
\be
\label{eq:Inm}
\eta &:= \int_{\mathcal S} \left\{ \zeta_{[\mu } N_{\nu]}  \left[
\psi^\nu \nabla^\mu \sigma
- \nabla^\mu \left( \frac{H' g^{\theta \theta}}{g^{vr}} \tau ^\nu \right)
\right] - \frac{1}{2} \zeta^\mu \nabla_\mu \log(g^{vr}) \right\}\, .
\ee
\Cref{eq:Smarr_general} is our proposal for a generalised Smarr's formula, valid for non-Killing horizons. 
As expected, it is a relation connecting the area of sections of the event horizon with the Komar mass and angular momentum defined on the same section.

The breakdown of circularity is evident from the fact that $\overline{\kappa}$ is the average over the horizon of a quantity which is not a constant.
Moreover, the quantity $\overline{\sigma}_H$ that multiplies the Komar angular momentum $J_H$ is generically not equal to the angular velocity of frame dragging --- more precisely, its average is not equal to the average of the angular velocity.
Finally, the presence of the term $\eta$ is a true novelty with respect to the usual formulation of Smarr's formula. 
Since, generically, it does not vanish and it is not connected to any conserved quantity of the spacetime, it is the thermodynamical analogue of a dissipative term --- see, e.g.  \cite{Pokrovski_2005}.
Non-circular black holes thus seem to behave as open systems. 
Even within this interpretation, $\eta$ as written in \cref{eq:Inm} does not lend itself to an immediate geometrical understanding. 
As we show below, considering minimal circularity breaking or introducing specific assumptions on the topology of $\mathcal S$ renders the treatment of $\eta$ somewhat more intuitive.

\subsubsection{\label{subsubseq:minimal_Smarr}Minimal case}

Several important simplifications take place if $H' = 0$.
First of all, one can easily show that in this case
\be \label{eq:Inm_minimal}
\eta = 0 
\quad
\left( H' = 0 \right) .
\ee
Indeed, the second term in the square bracket of \cref{eq:Inm} trivially vanishes, while the rest reads
\be
\eta = \frac{1}{2} \int_{\mathcal S} \left[\left( N_\nu \, \psi^\nu \right) \zeta^\mu \nabla_\mu \sigma -  \zeta^\mu \nabla_\mu \log(g^{vr}) \right]
\quad
\left( H' = 0 \right) ;
\ee
but, since $\zeta^\mu \propto \Xi^\mu$ when $H'=0$, $\zeta^\mu \nabla_\mu$ only contains derivatives in the Killing directions, along which both $\sigma$ and $g^{vr}$ are constant by construction.
Hence the whole integrand vanishes.

Moreover, in this case $\eval{\sigma}_{r=H} = \eval{\Omega}_{r=H}$, and therefore the average $\overline{\sigma}_H$ does coincide with the average of the angular velocity with which the horizon rotates.

Hence, in the minimal case $H'=0$, \cref{eq:Smarr_general} takes the much more familiar form
\be \label{eq:Smarr_minimal}
M_H=\frac{\overline{\kappa}}{4 \pi G}  \mathcal{A}_H + 2 \overline{\Omega}_H J_H   
\quad
\left( H' = 0 \right) \, ,
\ee
which only differs from the usual Smarr's formula in so far as the quantities $\overline{\kappa}$ and $\overline{\Omega}_H$ are averaged versions of the (non-constant) surface gravity and angular velocity.
At the risk of being pedantic, we wish to stress that $H'=0$ does not mean that the horizon is circular.

\subsubsection{\label{subsubsec:improved_Smarr}Improving the non-minimal case: the dominant convergence condition}

A significant improvement to the formula of \cref{eq:Smarr_general} can be achieved by imposing an additional constraint on the geometry.
Indeed, a classical result due to Hawking \cite{HawkingBlackHoles1972} ensures that, under certain assumptions, the topology of the sections $\mathcal{S}$ must be that of a two-dimensional sphere. 
If this is the case, then the tangent space of $\mathcal{S}$ must be spanned by $\tau^\mu$ and $\psi^\mu$, and therefore we must have
\be \label{eq:N_transverse}
N_\mu \tau^\mu=N_\mu \psi^\mu=0 \,.
\ee

The assumptions of Hawking's theorem are the following:
\begin{enumerate}
    \item stationarity and axisymmetry;
    \item asymptotic flatness; and
    \item the dominant convergence condition.
\end{enumerate}
Assumption 1 has been made throughout the whole article; 
the same holds for assumption 2;
assumption 3 is genuinely new and represents a (mild) loss of generality.
By \qu{dominant convergence condition}, we mean that for any two future-directed non-spacelike vectors $v^\mu$ and $w^\mu$, the following must hold:
\be \label{eq:DEC_R}
 \left( R_{\mu \nu} - \frac{1}{2}R g_{\mu \nu} \right) v^\mu w^\nu \ge 0\, .
\ee
This is tantamount to saying that $\left( R_{\mu \nu} - \frac{1}{2}R g_{\mu \nu} \right) v^\mu$ is not spacelike, and it implies the weak convergence condition.
Here we use the term \qu{convergence condition} instead of the more familiar \qu{energy condition}, since we wish to be agnostic on the form of the field equations; if these amount to a modified version of Einstein's equations, as is usually the case, we mean that the relevant energy condition holds for the effective stress--energy tensor that is equal, on shell, to the Einstein tensor.
For further details on energy conditions, and the violation thereof, see e.g.~\cite{VisserLorentzianWormholes1996,CurielPrimerEnergy2017,KontouEnergyConditions2020} and references therein.
Note that, in Hawking's proof, the dominant convergence condition is used to ensure the positivity of the intrinsic curvature of the sections $\mathcal{S}$.

Under these assumptions, \cref{eq:Inm} becomes
\be \label{eq:Inm_DEC}
\eta = \int_{\mathcal S} \left[  - \frac{H' g^{\theta \theta}}{g^{vr}} \zeta_{[\mu } N_{\nu]} \nabla^\mu \tau^\nu - \frac{1}{2} \zeta^\mu \nabla_\mu \log(g^{vr})\right] \,.
\ee
Incidentally, we point out that \cref{eq:N,eq:N_transverse} allow to determine the form of $N^\mu$ completely as
\be \label{eq:N_DEC}
\eval{N_\mu}_{r=H} = \eval{ \frac{g^{vv}}{2\alpha} \zeta_\mu - \delta^v_{\ \mu} }_{r=H}
\,, 
\ee
although this will not be needed for what follows.
What we will need, instead, is the following identity, which can be proven after some algebra and recalling that $\alpha \equiv 1$:
\be
 \label{eq:boh}
\zeta_{[\mu } N_{\nu]} \nabla^\mu \tau^\nu = N^\nu \tau^\mu \nabla_\mu \zeta_\nu  - \frac{1}{2} \tau^\mu \partial_\mu \log(g^{vr})\, .
\ee
Moreover, exploiting \cref{eq:zeta_decomp,eq:anti_nabla_zeta}, we can write
\be
- \frac{1}{2} \zeta^\mu \nabla_\mu \log(g^{vr}) &=  \frac{1}{2} \frac{H' g^{\theta\theta}}{g^{vr}} \tau^\mu \nabla_\mu \log(g^{vr}) \nonumber\\
&= \frac{H' g^{\theta\theta}}{g^{vr}} \left[ -\frac{N^\mu \tau^\nu}{2} \zeta_\mu \nabla_\nu \log(g^{vr}) \right] \nonumber\\
&= - \frac{H' g^{\theta\theta}}{g^{vr}} N^\nu \tau^\mu \nabla_{[ \mu} \zeta_{\nu]} \, .
\ee
Combining these results, we can finally write
\be
\eta = \int_\mathcal{S} \left[ - \frac{H' g^{\theta\theta}}{g^{vr}} N^\nu \tau^\mu \left( \nabla_\mu \zeta_\nu + 2 \nabla_{[ \mu} \zeta_{\nu]} \right) \right]\, .
\ee
As anticipated, this expression is somewhat simpler than \cref{eq:Inm}; yet, a full understanding --- beyond its interpretation as a dissipative term --- requires further investigations.

\subsection{\label{subsec:tunnel}Tunnelling and Hawking radiation}

Although inherently classical, (the generalised) Smarr's formula alludes to the existence of semiclassical effects also for non-circular horizons. 
As we will see in the next section, \cref{eq:Smarr_general} suggests that the identification of the horizon's area with the black hole entropy still holds for these kind of non-Killing horizons. 
This possibility gives a strong hint that black hole should evaporate in a similar way as the circular ones, in order not to give rise to thermodynamical paradoxes, as the violation of the generalised second law of black hole thermodynamics \cite{BekensteinBlackHoles1973,Bekenstein_Generalized_SecondLaw,Wald_GR,Liberatiprivcomm}.

Independently from that, the presence of modes which peel off exponentially from the horizon --- such as the ones studied in \cref{subsec:kpeel} --- is undoubtedly a smoking gun of Hawking radiation \cite{Hawking:1975vcx}. 
Mathematically speaking, while considering quantum fields living on a curved background, one can link the presence of a non-analytical term in the outgoing geodesic to a thermal character of the particle spectrum at infinity. 

Among the several ways in which Hawking radiation can be computed, here we will make use of the so-called \qu{tunnelling method}, which provides a quasi-local technique to describe the black hole's particle production and has shown to be extremely versatile in computing Hawking radiation in different scenarios. 
This method can be shown to be equivalent to the Bogolyubov calculation originally employed by Hawking \cite{Hawking:1975vcx} --- which maps the outgoing basis near the horizon into the basis for an observer at infinity --- since the temperature of the spectrum is uniquely determined \textit{at the local level} by the analytical structure of the modes near the horizon.

This kind of particle production effect has been proven to be extremely robust under the change of framework. 
Static and stationary horizons \cite{Parikh:1999mf,Shankaranarayanan:2000gb,Vanzo:2011wq} (also for regular black holes \cite{Alonso-Bardaji:2025qft}), as well as dynamical ones \cite{CriscienzoHawkingRadiation2007,Hayward:2008jq,Giavoni:2020gui}, Lorentz-violating horizons \cite{Berglund:2012fk,DelPorro:2022vqi,DelPorro:2023lbv,DelPorro:2025zyv}, and acoustic horizons \cite{DelPorro:2024tuw} are just a few examples where the tunnelling method has been applied. 
In all these cases Hawking radiation arises in very similar (yet different) fashions. 
Non-circular black holes make no exception, as we shall see in a moment.

Let us start by considering a test massless scalar field $\varphi$, whose dynamics is described by the Klein--Gordon equation
\be \label{eq:KG}
 \Box \varphi =0 \,.
\ee
Due to their exponential blueshift, the outgoing subset of solutions of \cref{eq:KG} is \textit{exactly} described by the Wentzel--Kramers--Brillouin (WKB) approximation of the field $\varphi$
\be \label{eq:WKB}
\varphi = \varphi_0 \, e^{i S} \,,
\ee
where the phase $S$ is called \qu{point-particle action} and $\varphi_0$ is a slowly varying amplitude. 
Indeed, if $S$ is a rapidly varying function --- as it happens near the horizon for the outgoing modes --- \cref{eq:KG} becomes
\be \label{eq:WKB_LO}
g^{\mu \nu} \nabla_\mu S \nabla_\nu S =0 \,.
\ee
Namely, the field $\varphi$ can be described by its eikonal approximation. 
Introducing the usual notation $k_\mu=\nabla_\mu S$, \cref{eq:WKB_LO} takes the form of a dispersion relation for a massless point particle with action $S$
\be 
k_\mu k^\mu=0 \,.
\ee
Its trajectory is the curve which minimises $S$, hence corresponding both to an outgoing null geodesic and to a constant phase contours of the WKB field $\varphi$, since, by definition
\be 
S= \int k_\mu {\rm d}x^\mu \Rightarrow 0 = \delta S = k_\mu \dot x^\mu = k_\mu k^\mu \,.
\ee

The shape of such trajectory has already been studied in \cref{subsec:kpeel}. 
As usual, thanks to the invariance under $\xi^\mu$ and $\psi^\mu$, we can label all the solutions of \cref{eq:KG} through their Killing energy $\omega$ and angular momentum $m$ via the eigenvalue equations
\be 
\xi^\mu \partial_\mu \varphi_{\omega m}= \partial_v\varphi_{\omega m}=-i \omega \varphi_{\omega m} \quad \mbox{and} \quad \psi^\mu \partial_\mu \varphi_{\omega m}= \partial_\phi \varphi_{\omega m}=i m \varphi_{\omega m} \,.
\ee
This corresponds to setting $k_v=k_\mu \xi^\mu= -\omega$ and $k_\phi=k_\mu \psi^\mu=m$, as done in \eqref{eq:KillingEL}. 
Hamilton's equations \eqref{eq:xdot} define the tangent vector to the constant phase contours of the action. 
In particular, near the horizon:
\be 
\frac{\dd v}{\dd z}= \frac{2 g^{vr}}{K_H} \frac{1}{z} + \mathcal O(1) \,, \qquad  \frac{\dd \phi}{\dd z}= \frac{2 g^{r \phi}}{K_H} \frac{1}{z} + \mathcal O(1)\,.
\ee
So, at the leading order, the constant-phase contours are expressed by the line element
\be \label{eq:classical_traj}
0= \eval{\frac{2 g^{vr} \omega - 2 g^{r \phi} m}{K}}_{r=H} \frac{\dd z}{z} - \omega \dd v + m \dd \phi\,,
\ee
or, equivalently
\be 
S= -\omega  v + m  \phi + \eval{\frac{2 g^{vr} \omega - 2 g^{r \phi} m}{K}}_{r=H} \log(z) \,.
\ee

As expected, the outgoing modes enjoy a logarithmic non-analyticity near the horizon, which is the key feature to extract the thermal properties of the horizon. Technically, the link is provided by the tunnelling rate formula
\be \label{eq:tunnelling_rate}
\Gamma=e^{ - 2 {\rm Im}( S) } \,.
\ee
Here, the imaginary part of the action $S$ comes from connecting the outgoing point particle trajectories on the two sides of the horizon with a classically forbidden complex path across the horizon itself. 
In other words, we integrate the right-hand side of \cref{eq:classical_traj} in the complex $z$ plane performing a small shift of the $z=0$ pole along the imaginary axis. 
The procedure is independent from the path of integration $(v(z), \, \phi(z), \, z)$, as long as the functions $v(z)$ and $\phi(z)$ are taken to be regular at $z=0$. 
So, chosen $z_1 <0<z_2$, we can simply integrate along $\dd v= \dd \phi=0$ to get
\be \label{eq:ImS}
{\rm Im}(S)= \lim_{\varepsilon \to 0^+} {\rm Im} \left[ \eval{\frac{2 g^{vr} \omega - 2 g^{r \phi} m}{K}}_{r=H} \int_{z_1}^{z_2}  \frac{\dd z}{z - i \varepsilon} \right]= \frac{\pi \left(\omega- m \sigma_H \right)}{\kp} \,,
\ee
where we used \cref{eq:k_peeling} in the last step in order to introduce $\kp$ and $\sigma_H$ is the usual function $\sigma=g^{r \phi}/g^{r v}$ evaluated on the horizon. Plugging this result into \cref{eq:tunnelling_rate} we get the tunnelling rate
\be \label{eq:tunnelling_rate_2}
\Gamma=\exp \left[ -\frac{2\pi}{\kp}\left(\omega- m \sigma_H \right) \right] \,.
\ee

If the quantum state for the field $\varphi$ is chosen to be the Unruh state (namely vacuum on $\mathscr{I}^-$ and on the horizon), the rate of \cref{eq:tunnelling_rate_2} describes a particle spectrum at infinity with a density
\be
\langle \hat{\mathcal N}_{\omega m} \rangle = \frac{\upsilon(\omega,m)}{\exp \left[ \frac{2\pi}{\kp}\left(\omega- m \sigma_H \right) \right]-1}
\ee
that describes a Bose--Einstein distribution with temperature $T_H$ and chemical potential $\mu_H$, respectively given by
\be
T_H= \frac{\kp}{2 \pi} \qquad \mbox{and} \qquad \mu_H=m \sigma_H \,,
\ee
while $\upsilon(\omega,m)$ is a grey-body factor. Once again, this result resonates with the well-known treatment of the circular case: the emission of the black hole is affected by the rotation and the co-rotating objects ($m \sigma_H>0$) are more likely to be emitted. However, while in the case of the Kerr geometry, the chemical potential is given by the constant $\mu_H^{\rm Kerr}= m \Omega_H $ and represents just a constant shift in energy, here $\mu_H$ depends on the position. In particular, in the minimal case we have $\sigma_H= \Omega_H$, as shown in \cref{eq:sigma_id}, and the chemical potential acquires the same shape as in the circular case, but with a non-constant $\Omega_H$.

A similar discussion can be done for the temperature: since $\kp=\kn$ depends on the point on the horizon, the resulting emission is anisotropic for both non-circular cases. This seems to point towards an out-of-thermal-equilibrium horizon.\footnote{Similar considerations were made by us in \cite{DelPorro:2025zyv}, in the case of a rotating Lorentz-violating black hole.} We shall comment about the physical consequences of this observation in the following sections.

Before concluding this section, let us comment on the role of $k_\theta$ when $H'=0$. In section \ref{subsec:kpeel} we computed the behaviour of $k_\theta$ as it approaches the horizon, finding \cref{eq:k_theta_minimal}. While apparently problematic, we show that this singularity is integrable since
\be
\int \log(z) \dd z= z \log(z) - z + {\rm const}
\ee
and finite in the limit $z \to 0$. Therefore, even though divergent, \cref{eq:k_theta_minimal} does not contribute to the imaginary part of the point particle action of \cref{eq:ImS} and to the black hole's emission rate of \cref{eq:tunnelling_rate}.

\subsubsection{Examples}
In order to connect more directly with \cite{BabichevPropertiesGeneral2025}, let us show explicitly the expression for the two examples of minimal and non-minimal breaking of circularity that we mentioned in section \ref{subsubsec:examples}. 

For the minimal case $H'=0$ and the metric we refer to is given by \cref{eq:minimal_eg}. The temperature is determined by $\kp=\kn$ of \cref{eq:minimal_kappa_eg}, while
\be
 \sigma_H=\eval{ \frac{a}{r^2 + a^2 + \delta}}_{r=r_+} =\Omega_H \,,
\ee
with $r_+= M + \sqrt{M^2-a^2}$. We immediately see that both the non-constancy of the temperature and the anisotropy of the emission are encoded in the $\theta$-dependence of $\delta$. 

For the non-minimal case, $H' \ne 0$, we consider the metric of \cref{eq:nonminimal_eg}. Again the temperature can be read off from \cref{eq:nonminimal_kappa_eg} since $\kp=\kn$. For the chemical potential we have
\be
\sigma_H=\eval{\frac{a}{r^2+a^2}}_{r=H} \neq \Omega_H \,,
\ee
where the non-constancy of $\sigma_H$ is given by the non-spherical shape of the horizon's section $r=H(\theta)= \eval{ \sqrt{- \tilde{ \Delta } } }_{r=H}$.

\subsection{\label{subsec:4law}On the four laws}

\Cref{subsec:Smarr,subsec:tunnel} suggest that stationary non-circular black holes could obey a modified version of the usual laws of black hole mechanics \cite{BardeenFourLaws1973}. The averaged version of Smarr's formula that we presented above indicates that these objects can still be characterised in terms of three extensive quantities, namely the mass $M_H$, the angular momentum $J_H$, and their horizon's surface area $\mathcal A_H$. At the same time, the non-constancy of the surface gravities and of the horizon's angular velocity are symptomatic of the lack of a global thermal equilibrium. This intuition is confirmed by the semiclassical analysis of the non-circular horizon's radiative properties, for which the emission turns out to be anisotropic and dependent on the position on the section $\mathcal S$. 

Hence, it is not clear at this point whether a treatment in terms of (global) equilibrium thermodynamics is truly viable, or whether a local notion of equilibrium could prove more insightful.
Nonetheless, it is worth gathering what we have understood so far and discussing how the four laws for black hole mechanics might appear in the present context.

\paragraph{Zeroth law}
For a Killing horizon, the zeroth law of mechanics states that the surface gravity is constant on the horizon. Since the surface gravity plays the role of temperature, this fact corresponds to the statement that a circular black hole is in thermal equilibrium.
In non-circular cases, this condition is violated. In their local definitions, all the notions of surface gravity depend on the position on the section $\mathcal S$.
Moreover, these notions differ, and it is not obvious if any of them is more insightful than the others --- although the fact that $\kp = \kn$ appears meaningful.
In the minimal case, the three definitions coincide and the surface gravity can be shown to remain constant along the generators' integral lines:
\be
\zeta^\mu \partial_\mu \kappa=0 \qquad (H'=0)\,,
\ee
but this does not hold on the transverse directions. So, we can conclude that, if a thermodynamical interpretation still holds, a non-circular black hole should be treated as a system out of thermal equilibrium, or perhaps one in \emph{local} thermal equilibrium.

\paragraph{First law}
Naively speaking, the first law appears nothing but a variational version of Smarr's formula.
We should point out, however, that such a law is properly formulated only in reference to solutions of some theory, specifically solutions that are \qu{infinitesimally close}, in some sense.
Hence, it is an inherently theory-dependent statement.
Assuming two such solutions exist, with masses, angular momenta, and horizon's areas ($M_H$, $J_H$, $\mathcal A_H$) and ($M_H+ \delta M$, $J_H+ \delta J$, $\mathcal A_H+ \delta \mathcal A$), respectively, in the circular case one can show that
\be
\delta M = \frac{\kappa}{8 \pi G} \delta \mathcal{A} + \Omega_H \delta J \,.
\ee
The derivation (see e.g.~\cite[ch.~5]{Poisson:2009pwt}) is extendible to non-Killing horizons. In the minimal case, the same proof gives exactly
\be \label{eq:firstLaw_M}
\delta M = \frac{\overline{\kappa}}{8 \pi G} \delta \mathcal{A} + \overline{\Omega}_H \delta J \qquad (H'=0) \,,
\ee
with an analogous interpretation of the quantities, which now appear in their averaged form. In the non-minimal case, the term $\eta$ generically appears as a possible source of energy dissipation.\footnote{Note that, while the minimal case necessarily entails $\eta = 0$ [cf.~\cref{eq:Inm_minimal}], it is not clear whether $\eta \neq 0$ \emph{necessarily} in the non-minimal case.} The natural extension to the case $H' \ne 0$ is
\be \label{eq:firstLaw_NM}
\delta M = \frac{\overline{\kappa}}{8 \pi G} \delta \mathcal{A} + \overline{\sigma}_H \delta J +  \delta \eta \,.
\ee
Therefore it seems possible to connect two different non-circular black holes with just a change in mass, while keeping the area and the spin fixed. In this sense, non-circular black holes seem to behave as open thermodynamical systems, where internal degrees of freedom can be used to dissipate energy \cite{Pokrovski_2005}. In terms of spacetime thermodynamics, this term might be read as the source of \qu{internal entropy production} of non-equilibrium configurations \cite{ElingNonequilibriumThermodynamics2006,Chirco_non_equilibrium_thermo}.

\paragraph{Second law}
The second law of black hole mechanics, known as \qu{area law}, states that the area of the event horizon cannot decease with time. The proof of this statement is due to Hawking \cite{HawkingBlackHoles1972,HawkingLargeScale1973}, and it is based on the same assumption as the theorem on the horizon's topology\footnote{To be precise, Hawking's proof is based on the \qu{null convergence condition}, according to which $R_{\mu \nu}l^\mu l^\nu \ge 0$ for all null vectors $l^\mu$. This condition, however, is implied by \cref{eq:DEC_R}, when the two vectors in the formula coincide with the same null vector $l^\mu$.}, which we recalled in \cref{subsubsec:improved_Smarr}. So, within the assumption of \cref{subsubsec:improved_Smarr}, we have that
\be
\delta \mathcal A \ge 0 \,,
\ee
which corroborates the thermodynamics interpretation of $\mathcal A_H$ as entropy. At the semiclassical level, the tunnelling calculation of \cref{subsec:tunnel} shows that these black holes evaporate, with a temperature and a chemical potential coherent with \cref{eq:firstLaw_M,eq:firstLaw_NM}. In this case, the energy conditions are violated semiclassically and the black hole obeys the generalised second law \cite{Bekenstein_Generalized_SecondLaw,Wald_GR}
\be
\delta S_{\rm tot}=\delta S_{\rm BH} + \delta S_{\rm ext} \ge 0 \,.
\ee
Here, the sum of the black hole entropy and its external contribution $S_{\rm ext}$ must increase.

\paragraph{Third law}
The third law states the impossibility of reaching extremal  (zero-temperature) configurations by passing through a finite sequence of equilibrium states. The classical proof is due to Israel \cite{Israel_1986_ThirdLaw}, and it involves the introduction of apparent horizons and their evolution in time. Although at first sight there seems to be no obstruction in extending the proof to non-circular black holes, a rigorous formulation has to be fully developed. Moreover, we point out that, even in the familiar realm of general relativity, the status of the third law is somewhat debated --- see e.g.~\cite{Kehle_2022_ThirdLaw,DafermosStabilityProblem2025} and references therein.

\section{\label{sec:concl}Discussion}
In this paper, we presented a first study of the mechanics of non-Killing horizons. In particular, we focused on a class of stationary and axisymmetric spacetimes for which the circularity conditions break down. These conditions, which are trivially satisfied in vacuum solutions of general relativity, are necessary and sufficient to ensure that a black hole's event horizon is also a Killing horizon. We showed this explicitly in \cref{subsec:circiffK}, where we highlighted that these conditions need only hold on the horizon itself in order for the conclusion to hold.

The non-Killing character of an horizon has a direct impact on its mechanical properties. In particular, basic geometrical notions, such as that of surface gravity, must be profoundly rethought. In \cref{sec:surfgrav}, we proposed a way to define different realisations of the surface gravity --- namely: inaffinity, normal, and peeling --- employing a hypersurface-orthogonal vector $\zeta^\mu$, which also coincides with the generator of the horizon. We showed that, when circularity is restored, the usual surface gravities are recovered. Although useful, these definitions suffer from an ambiguity, coming from the normalization of such vector. The same arbitrariness arises in the circular case, and it is addressed non-locally by fixing the normalisation of the Killing vector generating the horizon at infinity. As for the present case, this ambiguity is not completely resolved, since our study focuses on the physics local to the horizon. The choice we made is still geometrically motivated, but we point out that a different normalization would impact our results in a non-trivial way. 

A special remark is reserved to the definition of the \emph{peeling} surface gravity. This parametrises the way in which outgoing future-directed geodesics peel off of the horizon. As opposed to the Kerr geometry, non-circular spacetimes lack complete integrability and the study of geodesics is highly non-trivial. Yet, in \cref{subsec:kpeel}, we showed that defining the peeling surface gravity does not require a complete determination of the outgoing geodesics. The final, expected outcome of this analysis is that of a surface gravity which is not constant on the black hole's horizon, as a consequence of its non-Killing nature.

The first half of the paper served as a preparation towards a more concrete analysis on the mechanics of these objects. In \cref{sec:thermo}, we presented a derivation of the so-called Smarr's formula. Since our spacetimes still enjoy stationarity and axisymmetry, it is still possible to identify two conserved quantities associated to said symmetries, namely the Komar mass $M_H$ and the Komar angular momentum $J_H$. We showed that these two charges enter Smarr's formula along with the horizon's surface area $\mathcal A_H$ in a way that is very similar to the circular case. As an integral equation, the non-circular version of such a formula involves quantities averaged on the two-dimensional sections of the horizon. Specifically, our formula contains the average of the surface gravity and of the angular velocity.

Quite interestingly, we showed that, if circularity is broken \qu{minimally} --- in which case the horizon's sections are still spherical and the generator $\zeta^\mu$ is a combination of Killing vectors --- the Smarr's formula holds exactly as in the circular case, apart from the aforementioned averaging. Conversely, if circularity is broken \qu{non-minimally}, the non-circular version of Smarr's formula acquires an extra term, which we interpreted as a dissipative term, since it is not associated to any conserved quantity. The appearance of dissipation is perhaps one of the most interesting features of non-Killing horizons and definitely requires further analysis.

At the local level, a hint towards a thermodynamical interpretation is given by the horizon's radiative properties. The presence of geodesics which peel off of the horizon exponentially allows for particle production phenomena \textit{à la} Hawking. Through the tunnelling method, we showed that non-circular horizons emit in a way that is very similar to their circular counterparts, with the only difference that the emission is anisotropic, due to the non-constant character of the temperature. In this respect, we note a clear similarity with previous works, where similar conclusions were drawn for horizons without Lorentz invariance \cite{DelPorro:2025zyv}.

The tunnelling calculation matches exactly the local expressions for temperature and chemical potential that appear in Smarr's formula and suggests that a consistent thermodynamical description might be possible. Indeed, in the final section of this paper, \cref{subsec:4law}, we discussed the status of the four laws of black hole mechanics in the non-circular case. Our analysis suggests that these objects may be considered as out of their thermal equilibrium (but not of their thermodynamical equilibrium) and a local thermodynamical treatment might be needed to better describe our result. Still, and quite remarkably, they seem to obey a generalised (and averaged) version of the first law, ensuing from Samarr's formula. The second law seems to hold as well, provided the assumptions of Hawking's area theorem are fulfilled. Those assumptions do not involve circularity, so our treatment is just as general as the familiar one for the circular case. In addition, the presence of Hawking's radiation points to a generalised second law. A precise formulation of the third law, which involves a study of apparent horizons, is beyond the scope of this work.

Evidently, to achieve a comprehensive understanding of non-Killing horizons, further investigation is required. As already mentioned, we believe that a local thermodynamical description might be more insightful for the physics of our case, both in describing the role of the dissipative term appearing in the non-minimal case, and in clarifying the role of the non-constant surface gravity. Engaging in speculation, one might wonder whether both effects point towards a relaxation of a non-circular horizon to a (circular) equilibrium state. In particular, the presence of an anisotropic temperature might induce tangential fluxes on the horizon, generating a backreaction on the geometry.

Finally, we emphasise that the lack of circularity appears to be a rather generic feature of stationary and axisymmetric solutions in theories of gravity beyond general relativity.
Hence, understanding the extent to which the familiar thermodynamic interpretation applies to non-Killing horizons is crucial.
We point out that such question is particularly timely, as current and future observations will probe gravity in its strong-field regime, and therefore allow to test general relativity and the Kerr hypothesis to unprecedented accuracy --- see e.g.~\cite{GW250114TestingHawking2025,GW250114Spectroscopy}.


\acknowledgments

We are particularly thankful to Stefano Liberati for his comments on the manuscript. FDP wants to thank Ariadna Ribes Metidieri for useful discussions. The research of FDP is supported by a research grant (VIL60819) from VILLUM FONDEN. The Center of Gravity is a Center of Excellence funded by the Danish National Research Foundation under grant No. 184.
JM acknowledges support of ANR grant StronG (ANR-22-CE31-0015-01).



\bibliographystyle{JHEP}
\bibliography{biblio.bib}

\end{document}